\documentclass{tlp}

\def\CORR{%
  \emph{To appear in Theory and Practice of Logic Programming (TPLP).}
  
}

\usepackage{aopmath}
\usepackage{listings}
\usepackage[pdftex]{graphicx}
\usepackage{hyperref}
\usepackage{picins}

\newcommand{\InsertImage}[2][1]%
   {\centerline{\fbox{{\includegraphics[width=#1\columnwidth]{#2}}}}}

\newcommand{\InsertImageNoBox}[2][1]%
   {\centerline{\includegraphics[width=#1\columnwidth]{#2}}}

\def\BL{\vspace{\baselineskip}}

\begin{document}

\bibliographystyle{acmtrans}

\long\def\comment#1{}

\def\BL{\vspace{\baselineskip}}

\def\gp{\textsf{GNU Prolog}}
\def\gplc{\texttt{gplc}}
\def\pl2wam{\texttt{pl2wam}}
\def\fd2c{\texttt{fd2c}}
\def\wamcc{\texttt{wamcc}}
\def\clpFD{\texttt{clp(FD)}}
\def\C#1{\multicolumn{1}{c|}{#1}}
\def\D#1{\multicolumn{1}{c||}{#1}}
\def\SlF{$\downarrow$~}
\def\Xinr{\mbox{\tt $X$~in~$r$}}


\def\Param#1{\texttt{\textit{#1}}}

\setlength{\leftmargini}{1.2em}
\setlength{\leftmargin}{1.5em}

\submitted{4th October 2009} 
\revised{1st March 2010} 
\accepted{22nd November 2010}

\title{On the Implementation of \gp{}}

\author[D.~Diaz, S.~Abreu and P.~Codognet]{
  DANIEL DIAZ \\
    University of Paris 1 \\
    90 rue~de Tolbiac    \\
    75013 Paris, FRANCE    \\
    E-mail: \texttt{Daniel.Diaz@univ-paris1.fr}
  \and %
  SALVADOR ABREU \\
  Universidade de \'Evora and CENTRIA FCT/UNL \\
    Largo dos Colegiais 2 \\
    7004-516 \'Evora, PORTUGAL \\
    E-mail: \texttt{spa@di.uevora.pt}
  \and %
  PHILIPPE CODOGNET \\
    JFLI, CNRS / University of Tokyo \\
    JAPAN \\
    E-mail: \texttt{Philippe.Codognet@lip6.fr}
}

\pagerange{\pageref{firstpage}--\pageref{lastpage}}
\volume{\textbf{10} (3):}
\jdate{January 2011}
\setcounter{page}{1}
\pubyear{2011}

\maketitle{}

\label{firstpage}

\begin{abstract}
  \CORR{}
  \gp{} is a general-purpose implementation of the Prolog language,
  which distinguishes itself from most other systems by being, above
  all else, a native-code compiler which produces standalone
  executables which don't rely on any byte-code emulator or
  meta-interpreter.  Other aspects which stand out include the
  explicit organization of the Prolog system as a multipass compiler,
  where intermediate representations are materialized, in Unix
  compiler tradition.  \gp{} also includes an extensible and
  high-performance finite domain constraint solver, integrated with
  the Prolog language but implemented using independent lower-level
  mechanisms.  This article discusses the main issues involved in
  designing and implementing \gp{}: requirements, system organization,
  performance and portability issues as well as its position with
  respect to other Prolog system implementations and the ISO
  standardization initiative.
\end{abstract}


\section{Introduction}
\label{sec:introduction}
\gp{}'s roots go back to the start of the 1990s at the Logic
Programming research team at INRIA Rocquencourt, in Paris.  Philippe
Codognet planned on implementing a low-level version of his
intelligent backtracking techniques and opted to do so on the
state-of-the art SICStus Prolog system, for which he obtained the
source code.  This task was handed down to Daniel Diaz, at that time
an M.Sc.~student.  However, at the time SICStus Prolog was already a
very large scale and complex system: version 2.1 had about 70,000 lines of
highly tuned and optimized C and Prolog code: clearly not the easiest
platform on which to carry out independent, low-level experiments.
So we took it upon ourselves to develop yet another implementation of
Prolog which would meet the following requirements:

\begin{itemize}

\item The system ought to serve as the basis for several
  research-oriented extensions such as: intelligent backtracking,
  co-routining (\texttt{freeze} was a hot topic back then),
  concurrency, constraints, etc.

\item The system would be made available freely to all researchers.

\item The system would be portable by design, not tied to any
  particular architecture.

\item The core of the system must be simple and lightweight, unlike
  SICStus.

\item The base system performance should be good.  The rationale being
  that a system designed to be extended needs to provide good base
  performance.  
  We targeted performance close to that of SICStus Prolog native code.

\end{itemize}
The last two points (simplicity and performance) were hard to
reconcile.  This was particularly true at a time where research on WAM
optimization was a hot topic: choice points, backtracking,
unification, indexing, register allocation, etc.: all aspects of the
WAM were the object of published research on optimizations thereof.
The goals we had set for ourselves seemed difficult to reach and
rather ambitious: performance with a simple implementation, all done
with little or no optimizations.

At the time, most Prolog systems were based on byte-code emulators,
written in C or assembly language. 
We decided we would need to compile to native code in order to recoup
the relative performance loss due to the inherent simplicity of the
WAM model we were to adopt.
It remained to be seen \emph{how} we would go about producing native
code.  At the time, producing native code seemed to be the best thing
one could possibly do: see BIM-Prolog, SICStus Prolog, Aquarius
Prolog, to name a few.  The SICStus approach was to retain its usual
emulated byte-code and only present the option of producing native
code for select architectures\footnote{Sparc under SunOS was the
  chosen one.}, which could be transparently mixed with emulated
predicates.  The approach followed by Peter van Roy with Aquarius
Prolog was different and more traditional: compilation was separate
from execution, as in regular programming
languages~\cite{DBLP:journals/computer/RoyD92}.  A ``command line''
compiler would translate a Prolog program into a native-code
executable.  Unfortunately, simplicity was apparently not a goal of
the exercise: Aquarius' abstract machine, the BAM, was lower-level and
finer-grained than the WAM and comprised more than 100 instructions,
some of which were specializations created on-the-fly by the compiler
using abstract interpretation techniques.  Other ``native code
generation'' approaches were surfacing, which would generate C rather
than an actual assembly language: these included Janus~\cite{Janus}
which translated Prolog to C,
KL/1~\cite{DBLP:conf/plilp/ChikayamaFS94} which compiled a language
different from Prolog (a committed-choice language, with don't-care
nondeterminism) or Erlang~\cite{DBLP:conf/iclp/Hausman93}, a
functional language only loosely related to Prolog.  A significant
difficulty that compilers to C had to deal with was the orthogonal
control dimension due to backtracking, which complicates stack frame
management beyond what C can normally do.

Our choice was to translate to C via the WAM, decorating the generated
code with direct assembly language instructions, to handle native
jumps which correspond to Prolog control transfers, such as the
\texttt{call}, \texttt{execute} or \texttt{proceed} instructions.  The
system we built was called
\texttt{wamcc}~\cite{DBLP:conf/iclp/CodognetD95} and most WAM
instructions were directly replaced by the equivalent C code, inlined
at compile-time via a set of C Preprocessor macros.  More complex
instructions result in library function calls, this was the case with
unification instructions, for example.  Lastly, \wamcc{} was the first
documented Prolog system to rely on the hardware (MMU) to detect stack
or heap overflows, by placing unmapped pages at the limits of the
dynamic memory areas: accessing these would raise an exception and
interrupt the normal flow of execution.  This approach resulted in a
clear performance gain when compared with bounds checks being
performed on every allocation and was innovative at the
time.\footnote{These days, this functionality is available
  off-the-shelf, in the form of the \texttt{libsigsegv} library, on
  \href{http://libsigsegv.sourceforge.net/}%
  {\texttt{http://libsigsegv.sourceforge.net/}}.}  Our choices clearly
paid off, as \wamcc{} got a 60\% performance gain w.r.t.~emulated
SICStus and about 30\% performance loss w.r.t.~SICStus native, the
agreed-upon references of the time.

Since then, \wamcc{} has been used as a teaching tool in several
universities and, outside INRIA, as the starting point for research
work, e.g.~\cite{DBLP:conf/padl/FerreiraD99}.  The next step ought to
have been the implementation of intelligent backtracking in \wamcc{}.
This was not to happen: our interest shifted to the blooming area of
Constraint Logic Programming~\cite{DBLP:conf/popl/JaffarL87}, which
was suddenly and for the first time enabling Logic Programming for
large-scale industrial applications.  There is little doubt that
Pascal Van Hentenryck's PhD thesis work and the implementation of a
Finite Domain (FD) solver in the CHIP
language~\cite{DBLP:conf/iclp/Hentenryck89} were instrumental in CLP's
success.  The CLP mechanisms underlying CHIP were touted as highly
optimized, totalling 50K lines of C code and shrouded in wraps of
secrecy -- a very useful and interesting black box.  It took a few
years for the work describing
\texttt{cc(FD)}~\cite{DBLP:journals/lncs/HentenryckSD94} to provide
hints as to how an effective CLP system could be implemented.  The
following move for \wamcc{} was clear: it would become \clpFD{} and
introduce a simple extension to the WAM to integrate FD
constraints~\cite{DBLP:conf/iclp/DiazC93,DBLP:journals/jlp/CodognetD96}.
\clpFD{} was about 4 times faster than CHIP.

The worst problem we had with \wamcc{} (and consequently \clpFD{}) was
the excessively long time it took GCC to compile the generated C code.
Even for moderately-sized programs, the time quickly became
overwhelming and even our attempts to banish most inlining in favour
of library function calls were insufficient to bring the times down to
acceptable levels.

On closer inspection, the C language was being used as a
machine-independent assembly language, which we had to fool in order
to do jumps.  This went against the regular operation of a C compiler,
which expects regular function entry and exit to be the norm: as a
consequence the compiler was trying to do its task over programs which
were really too devious and too large for it to properly cope.  As we
did not really need all that C could express, we started looking for
alternative languages to compile into.  One possibility which looked
very interesting was \texttt{C--}~\cite{DBLP:conf/ifl/JonesNO97}.
Unfortunately this system was not developed to the point where it
would be actually useful for our purpose: it remains bound to a
limited set of back-ends.\footnote{For a long time \texttt{C--} was
  restricted to 32-bit x86, and even though this situation has
  evolved, the set of target architectures is still smaller than what
  we have attained with the MA tool.}  We then set out to specify and
implement an intermediate language of our own.
Our ``Mini-Assembly'' (MA) language would have to meet very basic
requirements: to directly handle WAM control and be able to call C
functions would be sufficient.  This 
simplicity was meant to promote the easy porting to common
architectures.

At that time, the Prolog standardization effort was in full swing,
which lead to the emergence of the specification document known as ISO
Core 1~\cite{iso-part1}.  We then committed to develop a completely
new, standard-compliant implementation of Prolog, which would use the
MA intermediate language instead of C in the compiler pipeline.  So
was \gp{} born, under the code name \texttt{Calypso}.

The Prolog language was not very popular (euphemism alert!) outside
the research community, and in particular no implementation of the
language was present in the GNU organization catalogue, whereas other,
similarly exotic languages such as Scheme, were.  We took it upon
ourselves to defend our case with GNU in late 1998 and the first
official release of \gp{} saw the light of day in April 1999.  \gp{}
is presently directly available for most Linux distributions.

The remainder of this article is structured as follows:
Section~\ref{sec:compilation-scheme} discusses the structure of the
\gp{} compiler pipeline with fully fleshed-out examples.  In
Section~\ref{sec:fd-constraints} we present \gp{}'s Constraint Logic
Programming design and implementation.
Section~\ref{sec:gp-and-standards} tackles the positioning of \gp{} in
the Prolog landscape and its relation to the ISO standardization
initiative.  Finally, in Section~\ref{sec:concl} we draw conclusions
from the experience acquired over the last years and lay out possible
plans for further developments, some of which are actively being
pursued by the authors of the present article as well as other
researchers.


\section{Compilation Scheme}
\label{sec:compilation-scheme}
In this section we detail the compilation scheme adopted in \gp{}.  As
previously stated, the main design decision was to use a simple WAM
and to compensate for the lack of optimizations by producing native
code, thereby avoiding the overhead of an emulator.  The compilation
process is then the key point of \gp{}.  In {\wamcc}, \gp{}'s
ancestor, we produced native code via C: a Prolog file was translated
into a WAM file, itself translated to C and finally to object code, by
the C compiler.  This approach had a major drawback: the time needed
to compile the C file; even for medium-sized Prolog sources, the time
needed by the C compiler was quickly dominating the entire compilation
process.  In {\gp} we decided to sidestep the issue by not compiling
to C but, instead, to directly generate assembly code.
The direct translation from the WAM to assembly code turns out to be a
significant effort as a translator must be written for each target
architecture.  We simplified this problem by defining an intermediate
language called \emph{Mini-Assembly} or MA for short, which can be
viewed as a machine-independent assembly language well suited to
be the target of a WAM translator.  This language will be detailed
below, in sections~\ref{sec:comp-wam2ma} and~\ref{sec:comp-ma-inst}.

\begin{figure}[htb]
  \centering
  \includegraphics[width=0.9\columnwidth]{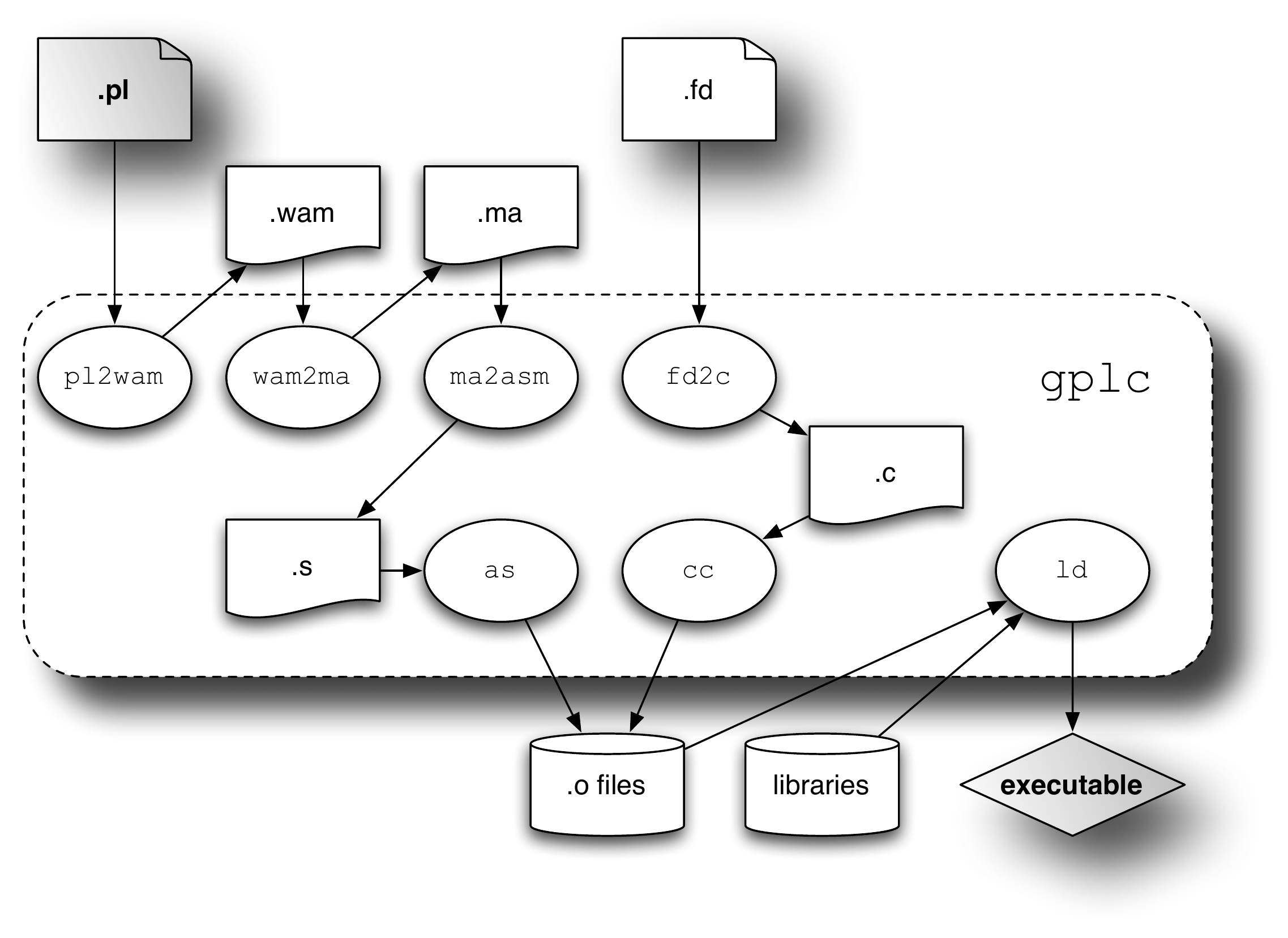}
  \caption{Compilation Scheme}
  \label{compil-scheme}
\end{figure}

One of the design goals for \gp{} was to offer a system which can
easily be extended by other research teams.  This requirement led us
to split the compiler into several passes, with distinct executables
for which the respective intermediate representations must be
materialized as plain text files or streams.  The passes are:

\begin{itemize}

\item[\texttt{pl2wam}] which compiles a Prolog source file into WAM
  code.

\item[\texttt{wam2ma}] which converts the WAM code to the MA language.

\item[\texttt{ma2asm}] that translates the abstract MA code to
  architecture-specific machine instructions: its output is an
  assembly language program.

\item[\texttt{fd2c}] compiles FD constraint definitions into C
  functions which perform constraint propagation at runtime.  See
  section~\ref{sec:fd-constraints} for more on this.

\end{itemize}
In addition to those \gp{}-specific compiler components, the
compilation process also involves the standard tools:

\begin{itemize}

\item[\texttt{as}] the assembler for the target architecture.

\item[\texttt{cc}] the C compiler

\item[\texttt{ld}] the link editor: to bind together all
  objects/libraries and provide a machine-dependent executable.

\end{itemize}
All of these are depicted in Figure~\ref{compil-scheme}. The general
compiler driver, called {\gplc}, manages all appropriate passes and
intermediate files, according to the provided input files and the
desired output. For instance, the user can mix input file types
(Prolog, WAM, MA, object files, libraries, etc.) as well as ask the
compiler to stop after any intermediate stage. The type of a file is
determined using the suffix of its file name and is used to select
its processor in the toolchain.

\subsection{Compiling Prolog to WAM}
\label{sec:comp-pl2wam}

Compiling Prolog to WAM is a well known and documented subject.  {\gp}
is based on the original WAM~\cite{Warren83,DBLP:books/mit/AitKaci91}
but uses a simple one-level indexing mechanism instead.  As previously
mentioned, the WAM variant we are compiling to is not very
sophisticated, for instance the code for structure unification does
not handle read and write modes separately, shallow backtracking is
not implemented, there is no separate choice stack (choice-points
reside in the local stack), to name but a few known techniques which
are \emph{not} used, for the sake of simplicity.  Nevertheless, a few
simple optimizations did get implemented, namely:
\ifx{1}{2}
\begin{itemize}

\item Improved register allocation.
\item Unification instruction reordering.
\item Inlining for selected simple built-in predicates.
\item Last subterm optimization.
\item Last call optimization.

\end{itemize}
\else %
improved register allocation, unification instruction reordering,
inlining for some built-in predicates, last call and last
subterm~\cite{Carlsson90} optimizations.

\fi
It is possible to disable any or all of these optimizations using
{\gplc} command line flags, a possibility which is particularly
interesting for people wishing to become familiar with the WAM
(e.g.~students taking declarative programming language implementation
courses.)

As an example let us consider a simple program with the canonical
``concatenate'' predicate.
\texttt{pl2wam} takes as input the source on the left of
figure~\ref{fig:prog:conc/3} and produces the file on the right.
\begin{figure}[htb]
\begin{tabular}{p{0.3\columnwidth}lp{0.5\columnwidth}}
\begin{lstlisting}
conc([], L, L).
conc([X|L1], L2, [X|L3]) :-
	conc(L1, L2, L3).
\end{lstlisting}
&
\hspace{3ex}
&
\begin{lstlisting}
file_name('/home/diaz/tmp/myprog.pl').

predicate(conc/3,1,static,private,user,[
    switch_on_term(1,2,fail,4,fail),
label(1),
    try_me_else(3),
label(2),
    get_nil(0),
    get_value(x(1),2),
    proceed,
label(3),
    trust_me_else_fail,
label(4),
    get_list(0),
    unify_variable(x(3)),
    unify_variable(x(0)),
    get_list(2),
    unify_value(x(3)),
    unify_variable(x(2)),
    execute(conc/3)]).
\end{lstlisting}
\end{tabular}
\caption{Predicate \texttt{conc/3} and its WAM representation.}
\label{fig:prog:conc/3}
\end{figure}
This code is very similar to that presented in~\cite{Warren83},
encoded as Prolog facts.  The fact for \texttt{file\_name/1} provides
the name of the Prolog source file which applies to the subsequent
predicate definitions.  Note that several instances of
\texttt{file\_name/1} may occur, as a result of \texttt{include}/1
directives.  The fact for \texttt{predicate/6} contains the code for
the predicate \texttt{conc/3}, as a list of WAM instructions.  Several
predicate properties are also stated here: \texttt{static} (as opposed
to \texttt{dynamic}), \texttt{private} (as opposed to \texttt{public})
and \texttt{user} (as opposed to built-in).

This code can be easily read and understood by humans (useful for
student use) and it can also directly serve as input for a Prolog
program like an emulator, a source-to-source optimizer or another
back-end as was done in the Prolog-to-EAM compiler, reported on
in~\cite{AndreAbreu:2010ciclops}.  The drawbacks of this choice are: a
not very compact representation (see the length of the instruction
names, for instance) and the need for a non-trivial parser for the
next stage which, in the case of \gp{}, is handled by \texttt{wam2ma}.

As previously mentioned, the \gp{} WAM is not very optimized.
Consider the clause: \texttt{p(a, X) :- q(a, X), r(X).}  Compiling it
results in the following sub-optimal code (2 WAM instructions could be
avoided):

\begin{lstlisting}
predicate(p/2,7,static,private,user,[
    allocate(1),
    get_atom(a,0),
    get_variable(y(0),1),
    put_atom(a,0),         % useless instruction !
    put_value(y(0),1),     % useless instruction !
    call(q/2),
    put_value(y(0),0),
    deallocate,
    execute(r/1)]).
\end{lstlisting}

\noindent
A WAM instruction cache could solve this as explained
in~\cite{Carlsson90}.  Basically, the cache remembers the current
values of the WAM registers.  When a \texttt{put} instruction occurs,
if the cache detects the wanted data is already present in a register
it replaces the \texttt{put} instruction by a register move
instruction (hoping the register optimizer will delete this move
intruction).  Optimizations such as this are not included in our
current compiler, which remains simple--about 3000 lines of Prolog
code--yet adequately efficient.  Consider file \texttt{bool.pl} and
the corresponding \texttt{bool.wam} shown in
figure~\ref{fig:prog:is_true/3}:
\begin{figure}[htb]
\begin{tabular}{p{0.3\columnwidth}lp{0.5\columnwidth}}
\begin{lstlisting}
is_true(true).

is_true(not(E)) :-
	is_true(E), !,
	fail.
is_true(not(_)).

is_true(and(E1, E2)) :-
	is_true(E1),
	is_true(E2).
\end{lstlisting}
& &
\begin{lstlisting}
file_name('/home/diaz/tmp/bool.pl').

predicate(is_true/1,1,static,private,user,[
    load_cut_level(1),
    switch_on_term(3,4,fail,fail,1),
label(1),
    switch_on_structure([(not/1,2),(and/2,10)]),
label(2),
    try(6),
    trust(8),
label(3),
    try_me_else(5),
label(4),
    get_atom(true,0),
    proceed,
label(5),
    retry_me_else(7),
label(6),
    allocate(1),
    get_structure(not/1,0),
    unify_variable(x(0)),
    get_variable(y(0),1),
    call(is_true/1),
    cut(y(0)),
    fail,
label(7),
    retry_me_else(9),
label(8),
    get_structure(not/1,0),
    unify_void(1),
    proceed,
label(9),
    trust_me_else_fail,
label(10),
    allocate(1),
    get_structure(and/2,0),
    unify_variable(x(0)),
    unify_variable(y(0)),
    call(is_true/1),
    put_value(y(0),0),
    deallocate,
    execute(is_true/1)]).
\end{lstlisting}
\end{tabular}
  \caption{Predicate \texttt{is\_true/1} Prolog and WAM code.}
  \label{fig:prog:is_true/3}
\end{figure}
the generated WAM code is, in this case, close to optimal.  Observe
how cut is handled: with the first instruction (line~4), the cut level
is stored in WAM temporary register \texttt{X(1)}.\footnote{In fact,
  this would actually be the first available \texttt{X} register,
  i.e.~\texttt{X(}arity\texttt{)} since we count from zero.  This
  technique is similar to that used in XSB~\cite{rao97xsb} or
  SICStus~\cite{Carlsson90}.}  Afterwards, it is treated as any other
Prolog variable: it only gets copied into a permanent
variable--\texttt{Y(0)} at \texttt{label(6)}--because its value is
needed after the first chunk, for the cut.

In short: \texttt{pl2wam} is a simple and portable Prolog-to-WAM
compiler, written in Prolog, which produces text files with a
representation of WAM programs.  The quality of the generated code,
while not outstanding, is adequate for our purpose.


\subsection{From the WAM to Mini-Assembly}
\label{sec:comp-wam2ma}

The next stage in the \gp{} compiler pipeline translates WAM
instructions to our MA intermediate language.  This language has been
designed specifically for the execution of Prolog based on our
experience with {\wamcc} when compiling down to C: in {\wamcc} most
WAM instructions finally ended up as a C function call which performed
the associated task.  Only a few instructions were inlined (via C
macros) because the size of the resulting code would have been
prohibitive for the available C compilers.  In fact, C was being used
as a sort of machine-independent assembler but:
\begin{enumerate}
\item the C compiler was unaware of the situation and spent a lot of
  time analyzing and optimizing the code and;
\item C is not truly an assembler and its control model is based on
  function definition and calls, making the efficient handling of
  Prolog backtracking very difficult.
\end{enumerate}
These problems drove us to design an intermediate representation, the
\emph{Mini-Assembly} with the following features:

\begin{itemize}

\item It handles the control of Prolog well: WAM instructions like
  \texttt{call}, \texttt{execute}, \texttt{return} and \texttt{fail}
  result in native jumps.

\item It can call a C function with a wide variety of arguments and
  can use the return value in several ways:
  \begin {itemize}
  \item To store it in memory or in a WAM register.
  \item To test its value and, to fail if zero (like \texttt{fail}).
  \item To branch to the address specified by the return value
    (e.g.~the address returned by \texttt{switch\_on\_term}.)
  \end{itemize}

\item It has a small instruction set (to facilitate the mapping to
  concrete machines) and only knows about a subset of the WAM, mainly
  that which is necessary for execution control.\footnote{For
    instance, the \texttt{fail} instruction needs to know about the
    \texttt{B} register and a displacement from it to get the
    alternative address to backtrack to.}

\item It is possible to declare scalar variables and arrays (only of
  type \texttt{long}).

\item It is possible to declare \emph{initializer} code, which will be
  automatically executed at run-time.  This issue is further discussed
  in section~\ref{sec:comp-ma-inst}.

\end{itemize}
The next section discusses and details the MA Instruction Set Architecture.

\subsection{The MA Instruction Set}
\label{sec:comp-ma-inst}

Here is a description of each MA instruction:

\begin{description}

\item [\texttt{pl\_jump} \Param{pl\_label}:] continue execution at the
  predicate whose symbol is \Param{pl\_label}.  This symbol is an identifier
  whose construction is explained in later on.  This corresponds to the WAM
  instruction \texttt{execute}.

\item [\texttt{pl\_call} \Param{pl\_label}:] continue execution at the
  predicate whose symbol is \Param{pl\_label}, after having set the
  continuation register \texttt{CP} to the address of the very next
  instruction.  This corresponds to the WAM instruction \texttt{call}.

\item [\texttt{pl\_ret}:] continue execution at the address given by
  the continuation pointer \texttt{CP}.  This corresponds to the WAM
  instruction \texttt{proceed}.

\item [\texttt{pl\_fail}:] continue execution at the address given by
  the last alternative, i.e.~the \texttt{ALT} cell of the last choice
  point, itself given by the WAM register \texttt{B}.  This
  corresponds to the WAM instruction \texttt{fail}.

\item [\texttt{jump} \Param{label}:] continue execution at the
  symbol \Param{label}, which can be any (predicate local) label.
  This instruction is used when translating WAM indexing instructions
  (e.g.~\texttt{try}, \texttt{retry} or \texttt{trust}) to perform
  local control transfer, i.e.~branching within the same predicate.
  This specialized version of \texttt{pl\_jump} exists because, in
  some architectures, local jumps can be optimized.

\item [\texttt{call\_c} \Param{f\_name}\texttt{(\Param{arg},...)}:]
  call the C function \Param{f\_name} with arguments \Param{arg},...
  Each argument can be an integer, a floating point number (C
  \texttt{double}), a string, the address of a label, the address or
  contents of a memory location, the address or contents of a WAM
  \texttt{X} register or \texttt{Y} permanent variable.  This
  instruction is used to translate most WAM instructions.  The return
  value of the function call can only be accessed by one of the
  \texttt{*\_ret} instructions which follow.

\item [\texttt{fail\_ret}:] perform a Prolog fail (like
  \texttt{pl\_fail}) only if the value returned by the preceding C
  function call is 0.  This instruction is used after a C function
  call returning a boolean to indicate its outcome, typically
  unification success.

\item [\texttt{jump\_ret}:] continue execution at the address returned
  by the preceding C function call.  This instruction makes it
  possible to use C functions to determine where to transfer control
  to.  For instance, the WAM indexing instruction
  \texttt{switch\_on\_term} is implemented by a C function
  returning the address of the selected code.

\item [\texttt{move\_ret} \Param{target}:] copy the value returned by
  the previous C function call into \Param{target} which can be either
  a memory location or a WAM \texttt{X} or \texttt{Y} variable.

\item [\texttt{c\_ret}:] C return.  This instruction is used at the
  end of the initialization function (see below) to give the control
  back to the caller.

\item [\texttt{move} \Param{reg1}\texttt{,} \Param{reg2}:] copy the
  contents of the WAM \texttt{X} or \texttt{Y} variable \Param{reg1}
  into \Param{reg2}.

\end{description}
The extreme simplicity of the MA language is noteworthy.  Observe,
however, the presence of the \texttt{move} instruction which performs
a copy operation on WAM \texttt{X} registers or on permanent
variables: while not strictly necessary,\footnote{Instead, we could
  easily invoke a C function to copy the data, using the
  \texttt{call\_c} instruction and an extra library function.} moves
between variables are very frequent and the invocation of a C function
would be costly in terms of execution time.  This reflects a tradeoff
between the minimality of the instruction set and acceptable
performance.  It would be possible to extend the instruction set
(e.g.~adding arithmetic instructions) but doing so would complicate
writing the architecture-specific back-ends with little expected gain.

In addition to the above instructions, MA also supports declaration
statements.  In what follows, the keyword \texttt{local} is used for a
local symbol (only visible within the current object file) while
\texttt{global} allows others to see that symbol.

\begin{description}

\item [\texttt{pl\_code global} \Param{pl\_label}:]
  define a Prolog predicate with name \Param{pl\_label}.  At present
  all predicates are tagged \texttt{global} (i.e.~visible by all other
  Prolog objects), but \texttt{local} will be used when implementing a
  module system.

\item [\texttt{c\_code local\textrm{/}global\textrm{/}initializer}
  \Param{label}:] define a function that can be called from C.  The
  use of \texttt{initializer} ensures that this function will be
  executed early, before the the Prolog engine is started.  At most
  one \texttt{initializer} function may be declared per file.

\item [\texttt{long local\textrm{/}global \Param{id}
    = \Param{value}}:] allocate the space for a \texttt{long} variable
  whose name is \Param{id} and initialize it with the
  integer \Param{value}.  The initialization is optional.

\item [\texttt{long local\textrm{/}global \Param{id}(\Param{Size})}:]
  allocate the space for an array of \Param{Size} \texttt{long}s whose
  name is \Param{id}.

\end{description}
The WAM to MA translation done by \texttt{wam2ma} is performed in
linear time w.r.t.~the size of the WAM file (the translation is
performed on the fly as the WAM file is being read).  This is the
behaviour that we sought in generating MA rather than C.

Like with the \texttt{pl2wam} phase, the result of \texttt{wam2ma} is
a text file that can be easily used as input for another program.

We now present the MA code obtained for our \texttt{bool.pl} example
(we have added the corresponding WAM instruction as comments):

\begin{lstlisting}
pl_code global X69735F74727565_1
  call_c   Pl_Load_Cut_Level(&X(1))              ; load_cut_level(1)
                                                 ; switch_on_term(3,4,fail,fail,1)
  call_c   Pl_Switch_On_Term_Var_Atm_Stc(&Lpred1_3,&Lpred1_4,&Lpred1_1)
  jump_ret   
Lpred1_1:
  call_c   Pl_Switch_On_Structure(st(0),2)       ; switch_on_structure(...)
  jump_ret   
Lpred1_2:	
  call_c   Pl_Create_Choice_Point2(&Lpred1_sub_0); try(6)
  jump     Lpred1_6
Lpred1_sub_0:	
  call_c   Pl_Delete_Choice_Point2()             ; trust(8)
  jump     Lpred1_8
Lpred1_3:
  call_c   Pl_Create_Choice_Point2(&Lpred1_5)    ; try_me_else(5)
Lpred1_4:
  call_c   Pl_Get_Atom_Tagged(ta(0),X(0))        ; get_atom(true,0)
  fail_ret   
  pl_ret                                         ; proceed 
Lpred1_5:
  call_c   Pl_Update_Choice_Point2(&Lpred1_7)    ; retry_me_else(7)
Lpred1_6:
  call_c   Pl_Allocate(1)                        ; allocate(1)
  call_c   Pl_Get_Structure_Tagged(fn(0),X(0))   ; get_structure(not/1,0)
  fail_ret   
  call_c   Pl_Unify_Variable()                   ; unify_variable(x(0))
  move_ret X(0)
  move     X(1),Y(0)                             ; get_variable(y(0),1)
  pl_call  X69735F74727565_1                     ; call(is_true/1)
  call_c   Pl_Cut(Y(0))                          ; cut(y(0))
  pl_fail                                        ; fail 
Lpred1_7:
  call_c   Pl_Update_Choice_Point2(&Lpred1_9)    ; retry_me_else(9)
Lpred1_8:
  call_c   Pl_Get_Structure_Tagged(fn(0),X(0))   ; get_structure(not/1,0)
  fail_ret   
  call_c   Pl_Unify_Void(1)                      ; unify_void(1)
  pl_ret                                         ; proceed
Lpred1_9:
  call_c   Pl_Delete_Choice_Point2()             ; trust_me_else_fail
Lpred1_10:
  call_c   Pl_Allocate(1)                        ; allocate(1)
  call_c   Pl_Get_Structure_Tagged(fn(1),X(0))   ; get_structure(and/2,0)
  fail_ret   
  call_c   Pl_Unify_Variable()                   ; unify_variable(x(0))
  move_ret X(0)
  call_c   Pl_Unify_Variable()                   ; unify_variable(y(0))
  move_ret Y(0)
  pl_call  X69735F74727565_1                     ; call(is_true/1)
  move     Y(0),X(0)                             ; put_value(y(0),0)
  call_c   Pl_Deallocate()                       ; deallocate
  pl_jump  X69735F74727565_1                     ; execute(is_true/1)

long local at(4)       ; array to store atoms
long local ta(1)       ; array to store tagged atom words
long local fn(2)       ; array to store tagged functor/arity words
long local st(1)       ; array to store switch tables

c_code  initializer Object_Initializer
  call_c   Pl_New_Object(&Prolog_Object_Initializer,
                         &System_Directives,&User_Directives)
  c_ret      

c_code  local Prolog_Object_Initializer
  call_c   Pl_Create_Atom("/home/diaz/tmp/bool.pl")
  move_ret at(0)
  call_c   Pl_Create_Atom("and")
  move_ret at(3)
  call_c   Pl_Create_Atom("is_true")
  move_ret at(1)
  call_c   Pl_Create_Atom("not")
  move_ret at(2)
  call_c   Pl_Create_Atom_Tagged("true")
  move_ret ta(0)
  call_c   Pl_Create_Functor_Arity_Tagged("and",2)
  move_ret fn(1)
  call_c   Pl_Create_Functor_Arity_Tagged("not",1)
  move_ret fn(0)
  call_c   Pl_Create_Pred(at(1),1,at(0),1,1,&X69735F74727565_1)
  call_c   Pl_Create_Swt_Table(2)
  move_ret st(0)
  call_c   Pl_Create_Swt_Stc_Element(st(0),2,at(2),1,&Lpred1_2)
  call_c   Pl_Create_Swt_Stc_Element(st(0),2,at(3),2,&Lpred1_10)
  c_ret      
c_code  local System_Directives
  c_ret      
c_code  local User_Directives
  c_ret      
\end{lstlisting}
From this example, one can observe that most WAM instructions map to a
C function call, following the subroutine-threading pattern.  As
previously mentioned, control instructions are directly translated to
their corresponding MA counterparts.  Another exception concerns moves
between WAM registers and permanent variables (e.g.~lines 29 and 51).
After further analysis of this example, several remarks can be made:

\begin{description}
\small

\item[Line 1:] \emph{(predicate label, \texttt{load\_cut\_level}...)}
  this is the start of the code associated to predicate
  \texttt{is\_true/1}.  As the MA language is to be mapped to the
  assembly language of an actual target machine, we decided that only
  ``classical'' identifiers can be used (a letter followed by letters,
  digits or the underscore character).  In particular, it is necessary
  to associate such an identifier to each predicate (referenced
  as \Param{pl\_label} in Section~\ref{sec:comp-ma-inst}).

  Since the syntax of assembly identifiers is more restrictive than
  the syntax of Prolog atoms (which may include any character using
  quotes) {\gp} uses a normalized hexadecimal-based representation for
  identifiers, where each predicate name is translated into a symbol
  beginning with an \texttt{X}, followed by the hexadecimal notation
  of the code of each character in the name, followed by an underscore
  and the arity.  For example, predicate symbol \texttt{is\_true/1} is
  encoded as the symbol \texttt{X69735F74727565\_1} (\texttt{69} is
  the hexadecimal representation of ``\texttt{i}'', \texttt{73} of
  ``\texttt{s}'', and so on).

  The linker is responsible for resolving external references
  (e.g.~calls to built-in or user predicates defined in another
  object).  The output of the linker is filtered by {\gp} to
  descramble hexadecimal symbol encodings, in case there are errors
  (e.g.~an undefined predicate, multiple definitions for a predicate).

\item[Line 4:] \emph{(\texttt{switch\_on\_term}...)}  the
  \texttt{switch\_on\_term} WAM instruction maps to a C call to a
  specialized function \texttt{Pl\_Switch\_On\_Term\_Var\_Atm\_Stc}
  which takes only three arguments (it checks if the first argument is
  a variable, an atom or a structure).  This is more efficient than
  calling the general C function with all 5 arguments (integer and
  list) as the call is cheaper (fewer arguments are passed) and faster
  (fewer cases to test).

\item[Line 7:] a \texttt{switch\_on\_structure} in the WAM code
  results in the creation of a switch table (done in the initializer
  part).  At execution time, this table is passed to the C function
  (the other argument is the size of the table -- in this case, two
  elements: \texttt{not/1}, \texttt{and/2}).  A hash-table with only
  two elements is not very efficient, a nested if would be better:
  clearly, there is room for improvement.

\item[Lines 10-16:] functions handling choice points are also
  specialized (here the functions are for a choice point with two
  arguments: the first argument in \texttt{X(0)} and the cut-level in
  \texttt{X(1)}.  Such specialized functions exist for arity $<4$.
  For greater arities, the arity must be passed as a parameter to the
  generic function.

\item[Line 18:] the \texttt{get\_atom} WAM instruction maps to a
  C function call to \texttt{Pl\_Get\_\-Atom\_Tagged}.  This function
  takes as first argument a tagged atom, i.e.~a WAM word.  This value
  is created by the initializer function and stored in module-local
  array \texttt{ta} at index zero.\footnote{\texttt{ta} stands for
    ``tagged atom''.}  Doing so avoids having to call the tag/untag
  function at runtime.  Here it is used to dereference \texttt{X(0)}
  and unify its value with \texttt{ta(0)}.

  The same occurs on lines 25, 36 and 44 where the
  \texttt{get\_structure} instructions get mapped to calls to
  \texttt{Pl\_Get\_Structure\_Tagged} which takes a single-word
  encoding of the functor and arity.  These are created by the
  initializer and stored in a module-local array \texttt{fn(...)}.

\item[Lines 55-58:] several arrays are declared to store atoms, tagged
  atoms, tagged functor/arity and switch tables.

\item[Line 60:] the initializer is declared.  It simply calls a C
  function to register this new object (an object file generated by
  the compilation process).  It passes 3 function pointers:
  \begin{itemize}
  \item \texttt{Prolog\_Object\_Initializer}: the initializer function
    for the object.  This function creates the atoms, the switch
    tables, the tagged words, etc.
  \item \texttt{System\_Directives}: executes system directives, such
    as \texttt{op/3} or \texttt{char\_conversion/2}.
  \item \texttt{User\_Directives}: is the entry point for the
    procedure which executes the user startup code (defined with
    Prolog \texttt{initialization} directives).
  \end{itemize}
  It is worth noting that the code needed to install the object
  (i.e.~the code in the body of \texttt{Prolog\_Object\_Initializer})
  cannot be directly executed in the initializer (i.e.~in
  \texttt{Object\_Initializer}) because that code is executed very
  early: when the OS~loads and runs the executable, i.e.~\emph{before}
  the \texttt{main} function is called.  At this time, the required
  global Prolog data structures (e.g.~atom hash table) are not yet
  allocated.  We therefore limit ourselves to registering the object
  and, only when all Prolog data areas are ready do the Prolog
  initializer functions get executed.

\item[Line 64-84:] the object initialization function creates the
  objects required by the code: atoms, tagged atoms, tagged
  functor/arity words, switch tables and stores these in the object's
  local arrays.  Atoms are classically hashed and thus can only be
  known at run-time (since we can have multiple files linked
  together). The initializer registers all needed atoms and stores
  them in local variables (e.g. in the \texttt{ta} array).  Notice
  that this could be optimized since once this information is created
  it remains constant during the execution of the program.  One could
  imagine a 2-pass optimizer: only execute all initialization
  functions to detect the values of all involved atoms, then recompile
  the whole using integer constants instead of MA array variables. The
  impact of this optimization would be very important if atoms are
  very used since it is much faster to pass an integer than loading it
  from the memory. Another benefit of this optimization would be the
  reduction of the startup time in applications which have a large
  number of atoms.  Finally, note that the initializer also registers
  the predicate \texttt{is\_true/1} with the \texttt{Pl\_Create\_Pred}
  function (line 79): this is only needed for meta-calls to work,
  because we need to associate an address (given here by the
  linker-resolved symbol \texttt{X69735F74727565\_1}) to the predicate
  symbol.  Other arguments are the file name and the line number where
  it is defined, and a mask containing the properties of the
  predicate.

\end{description}
To summarize, the Mini-Assembly language has a few control-flow
instructions, pseudo-ops to control constants and data areas, the C
function invocation operation and register movement instructions.
Predicate names are hashed into linker-friendly names.  These features
make it sufficient as the target for compiling WAM code.

\subsection{From Mini-Assembly to Actual Assembly}

The next stage consists of mapping the MA language generated in the
previous section to the target machine's actual instructions. Since MA
is based on a very small instruction set, the writing of such a
translator is inherently simple. However, producing machine
instructions is not an easy task. The first MA-to-assembly language
mapper was written with the help of snippets taken from a C file
produced by {\wamcc}: indeed, compiling a Prolog file to assembly by
means of \texttt{gcc} gave us a starting point for the translation, as
the MA instructions correspond to a subset of the C code.  We then
generalized this approach by defining a C file, each portion of which
corresponds to an MA instruction: the study of the assembly code
produced by \texttt{gcc} was our reference.  This provided preliminary
information about register use conventions, C calling conventions,
etc.  However, in order to complete the assembly code generator, we
need to refer to the technical documentation of the processor together
with the ABI (Application Binary Interface) used by the operating
system.

We now show portions of the assembly code for the previous example,
using the \texttt{linux/i86\_64} target.\footnote{We are still using
  the same example, meaning this is file \texttt{bool.s}.}  We focus
on the code for the last clause of \texttt{is\_true/1}, annotated with
the corresponding WAM \& MA code:

\begin{lstlisting}
fail:                            # fail (WAM inst)
   jmp     *-8(%r14)             #   pl_fail (MA inst)

X69735F74727565_1:               # predicate is_true/1
   ...
Lpred1_10:
                                 # allocate(1)
   movq     $1,%rdi              #   call_c Pl_Allocate(1)
                                 # get_structure(and/2,0)
   call     Pl_Allocate          #   call_c Pl_Get_Structure_Tagged(fn(1),X(0))
   movq     fn+8(%rip),%rdi
   movq     0(%r12),%rsi
   call     Pl_Get_Structure_Tagged
                                 #   fail_ret   
   test     %rax,%rax
   je       fail
                                 # unify_variable(x(0))
   call     Pl_Unify_Variable    #   call_c Pl_Unify_Variable()
   movq     %rax,0(%r12)         #   move_ret X(0)
                                 # unify_variable(y(0))
   call     Pl_Unify_Variable    #   call_c Pl_Unify_Variable()
   movq     2064(%r12),%rbx      #   move_ret Y(0)
   movq     %rax,-32(%rbx)
                                 # call(is_true/1)
   movq     $.Lcont1,2056(%r12)  #   pl_call X69735F74727565_1
   jmp      X69735F74727565_1
.Lcont1:
                                 # put_value(y(0),0)
   movq     2064(%r12),%rbx      #   move Y(0),X(0)
   movq     -32(%rbx),%rdx
   movq     %rdx,0(%r12)
                                 # deallocate,
   call     Pl_Deallocate        #   call_c Pl_Deallocate()
                                 # execute(is_true/1)
   jmp      X69735F74727565_1    #   pl_jump X69735F74727565_1


Object_Initializer:
   pushq    %rbx
   subq     $256,%rsp
   movq     $Prolog_Object_Initializer+0,%rdi  # call_c Pl_New_Object(...)
   movq     $System_Directives+0,%rsi
   movq     $User_Directives+0,%rdx
   call     Pl_New_Object
   addq     $256,%rsp            #   c_ret      
   popq     %rbx
   ret     

   .section .ctors,"aw",@progbits
   .quad    Object_Initializer
\end{lstlisting} 
\noindent A few observations:
\begin{description}
\small

\item[Line 1:] a label is defined to perform a WAM \texttt{fail}.
  Each time a fail is needed, a jump is performed to this label
  (e.g.~Line 16 for the MA instruction \texttt{fail\_ret}).  We can
  see that the last choice point frame (\texttt{B}) is stored in the
  x86\_64 register \%r14 and the alternative (\texttt{ALTB}) is the
  first 64 bits cell just below the address pointed by \texttt{B}.  An
  indirect jump does the work.


\item[Line 12-13:] a call to a C function yields an assembly
  \texttt{call} instruction, respecting the x86\_64 ABI: the arguments
  are passed via the registers \%rdi, \%rsi, \%rdx,...  Line 12 also
  reveals that the address of the bank of WAM temporaries
  (\texttt{X(...)}  variables) is kept in the register \%r12.  Other
  used registers are \%r13 for the top of the trail (\texttt{TR}) and
  \%r15 for the top of the heap (\texttt{H}).

\item[Lines 25-26:] a WAM \texttt{call} instruction gives produces 2
  assembly instructions.  The first one stores the next address (a
  local label) in the \texttt{CP} register (accessed as an offset from
  \%r12).  The second instruction simply jumps to the called
  predicate.

\item[Lines 38-47:] the initializer which calls a C function to
  register this object.

\item[Lines 49-50:] fill the CTORS sections (``constructors'') with a
  new entry: \texttt{Object\_\-Initializer}.  At run-time, the
  contents of this section is interpreted as an array of addresses,
  all of which are executed as functions (see
  section~\ref{sec:link-phase}).
\end{description}
Assembling \texttt{bool.s}
produces a relocatable object file called \texttt{bool.o} which can be
linked with the Prolog library and other modules to form an executable
image file.

\subsection{The Link Phase}
\label{sec:link-phase}

The last stage consists of linking all objects resuling from Prolog
sources (as explained above) with objects stemming from other sources
(e.g.~foreign C code), the {\gp} libraries and other objects (system
or third-party libraries.)  One design goal was to rely on standard
compiler tools to retain only what is necessary: the linker
(\texttt{ld} under Unix) links to an object library, from which only
the required modules are taken, thereby keeping the size of the final
executable down.  Since a Prolog source results in a classical object
file, several objects can be grouped in a library (e.g.  using
\texttt{ar} under Unix).  The Prolog and FD built-in libraries are
created in this way (and users can also define their own libraries).
Defining a library allows the linker to extract only the object files
that are necessary (i.e.~those containing referenced functions/data).
For this reason, {\gp} can generate small executables by avoiding the
inclusion of most unused built-in predicates.  On the other hand, the
linker cannot guess which built-in predicates will be called by a
meta-call.  To deal with this problem, \gp{} provides the directive
\texttt{ensure\_linked} which guarantees that a given predicate will
be linked (in fact, all it does is to create a simple reference to the
predicate in the assembly file, to force the linker to pull the
desired predicate in from the library).

As previously stated, each linked object includes initialization code
in which various housekeeping functions are performed.  This function
gets executed 
before any compiled Prolog code.
The ELF format allows the specification of global Object-Oriented
constructor code, which gets executed at the start and is collected
from several object modules.  We use this mechanism to initialize
\gp{} objects.

\subsection{\gp{} Executable Behaviour}
\label{sec:gp-exec-behav}

From the user point of view, the behaviour of an executable produced
by \gp{} consists of executing all \texttt{intialization/1}
directives.  If several \texttt{initialization/1} directives appear in
the same file they are executed in the order of appearance.  If
several \texttt{initialization/1} directives appear in different
Prolog files (i.e.~in different objects) the order in which they are
executed is implementation-defined.  However, on most machines the
order will turn out to be the reverse of the order in which the
associated files have been linked.  The traditional Prolog interactive
top-level interpreter is optionally linked with the rest of the
executable.  Should it be present, it gets executed after all the
other \texttt{initialization/1} directives have finished.  This
default behaviour is provided as a \texttt{main} function defined in
the \gp{} library.  So in the absence of a user-defined \texttt{main}
function the default function is executed.  Here is its definition:

\begin{lstlisting}
int main (int argc, char *argv[]) {
  int nb_user_directive;
  Bool top_level;

  nb_user_directive = Pl_Start_Prolog(argc, argv);
  top_level = Pl_Try_Execute_Top_Level();
  Pl_Stop_Prolog();

  if (top_level || nb_user_directive)
    return 0;

  fprintf(stderr, NOINITGOAL);
  return 1;
}
\end{lstlisting}
\begin{description}
\small

\item[Line 5:] the \texttt{Pl\_Start\_Prolog} allocates all data areas
  (stacks, tables, etc.) and, for each registered object, in reverse
  order, invokes its \texttt{Prolog\_Object\_Initializer},
  \texttt{System\_Dir\-ectives} and \texttt{User\_Directives}.  It
  returns the number of directives actually executed.

\item[Line 6:] if the top-level is linked then execute it.

\item[Line 7:] free all allocated areas.

\item[Lines 12-13:] warn the user that nothing has been executed,
  i.e.~there is no user directive and the top-level is not present in
  the executable.
\end{description}
The user can provide another \texttt{main} function, to customize this
behaviour.

\subsection{Bootstrapping the System}

Being written in Prolog, \gp{} -- and the \texttt{pl2wam} compiler in
particular -- relies on its own availability in order to recompile
itself.  We now discuss some aspects of the bootstrap process.

The parts of the Prolog compiler written in Prolog\footnote{Actually,
  the entire compiler is written in Prolog.} are expected to be
compilable by \gp{}. As a consequence, the \texttt{.pl} source files
need to be compiled using the \gp{} compiler pipeline, as described in
figure~\ref{compil-scheme}: in particular, there will have to be a
\texttt{.wam} file for each \texttt{.pl} source. These files are then
further compiled by the non-Prolog parts of the system
(\texttt{wam2ma}, \texttt{ma2asm}, the assembler and link editor.)

Note that a running Prolog system is only needed to get to the
\texttt{.wam} representation: from that point on, all compiler passes
are implemented as C programs.
In order to bootstrap \gp{} on a particular machine, one does not
actually need any working Prolog compiler, as the \texttt{.wam} files
for the \texttt{pl2wam} executable are provided with the source.

Another aspect worth mentioning involves quality assurance for the
Prolog-to-WAM compiler: a new version of the compiler should hit a
fixpoint for the contents of the \texttt{.wam} files: the files
produced by the compiler should converge to be identical to those
which make up the compiler itself.  The initial \texttt{.wam} files
may be produced from the Prolog sources either by \gp{} or by another
Prolog system.  The integrity of the generated Prolog compiler is
automatically verified by comparing the resulting \texttt{.wam} files
to the ones originally provided.

\subsection{Different Code Representations}
\label{code-rep}

The primary goal of \gp{} is to compile to native code and thus to
provide standalone executables, in the sense that references within
the program are statically resolved by the linker and the code is
directly available for execution. Emulator-based systems appear to
provide similar functionality by bundling the program bytecodes with
an emulator. However, \gp{} has also to handle dynamic Prolog clauses.
Generally speaking, \gp{} simply has to be able to meta-interpret.
This is the case when the programmer uses the \texttt{asserta} or
\texttt{assertz} built-in predicates: the clause will be stored and
(meta-)interpreted. The compiler will try to do a bit better than
this, in some situations: suppose, for instance, that a Prolog source
file contains a \texttt{:-~dynamic} directive for some predicate: the
native code for all defined clauses is generated. At run-time, even
though the predicate is dynamic, it's the native code that gets
executed: this ceases being so when the clause is removed with
\texttt{retract}/1. Clauses added at run-time will only be
meta-interpreted, i.e.~they will have no native code counterpart. In
the case of dynamic predicates, in addition to the native code, the
compiler also emits a system directive which records the term
associated to the clause (to be inspected using \texttt{clause/2}).

It can be argued that standardization efforts could have
differentiated among the two situations: to have one ``assert for code
only'' and another for ``data only.''

Out of respect for Prolog tradition, \gp{} also offers an interactive
top-level.  A major problem \gp{} has to face is the implementation of
the (in)famous \texttt{consult(FILE)} and \texttt{reconsult(FILE)}
predicates.  Several possibilities exist in a native-code system:

\begin{enumerate}
\item Read \texttt{FILE}, assert the code and have it
  meta-interpreted.
\item Compile \texttt{FILE} to byte-codes, which will be interpreted
  by a WAM byte-code emulator.
\item Compile \texttt{FILE} to native code and find a solution to
  dynamically load it into the running process.
\end{enumerate}
The first solution is simple to implement but obviously not very
efficient.  For the time being, we settled for the second solution: we
have developed a simple emulator to execute a binary representation
of the code provided by \texttt{pl2wam}.  This emulator is not
optimized at all but provides a speedup of about 3 when compared to
meta-interpreted code.

We plan on moving to the third approach, which is becoming feasible in
a portable way by resorting to native shared libraries, which can be
dynamically loaded or released from the running process memory.
Following this route frees us from having to use the byte-code
interpreter.  On the downside, the production of native code that can
be dynamically loaded is a bit more demanding because the machine code
has to be position-independent, which requires rewriting of the
architecture-specific back-ends for \texttt{ma2asm}.

To summarize, \gp{} currently manages three kinds of code:
\begin{itemize}%
\item interpreted code for meta-call and dynamically asserted clauses
\item emulated (byte-)code for consulted predicates
\item native code for statically compiled predicates
\end{itemize}
As a result, these three ways of representing and executing Prolog
programs need to be integrated, which turned out to be a demanding
requirement, as these models differ quite a bit. \gp{} has by no means
the exclusivity as far as this aspect is concerned: other Prolog
systems need to represent programs in more than two ways (BIM-Prolog
and SICStus for instance.)

\subsection{Discussion}
\label{sec:discussion}

This concludes the presentation of the \gp{} compilation scheme. Some
goals or aspects of the system are comparable to other systems, for
instance the native code implementation for SICStus Prolog
of~\cite{Haygood94} or Aquarius~\cite{DBLP:journals/computer/RoyD92}
which also aim at compiling Prolog to interpreter-less native code for
real architectures.

With respect to the direct generation of native code as opposed to
going through C, the latter has the advantage that it is easier to set
up, more portable and maintainable.  The downsides include high
compilation times (as a result of using a general-purpose, optimizing
C compiler), relatively low performance when generating standard C
code 
and, should one strive to improve performance by using nonstandard
extensions to the C language, the maintenance effort of the C compiler
itself.  Our option of direct native code generation benefits from
much better compilation times and potentially very high performance at
runtime.  The main drawback of this approach is its maintainability:
new targets must be explicitly programmed and adding new cross-cutting
features to the language or model requires an adaptation of all the
back-ends (e.g.~threads or dynamic linking.)

Native code generation is usually aimed at high performance. The
potential is high: absolute control over hardware register
usage,\footnote{Being able to use hardware registers favours a
  register-based abstract machine such as the WAM, as opposed to other
  approaches.} optimal tagging schemes, precise control flow, to name
but a few aspects.  However, in order to tap into this potential, rich
intermediate representations need to be devised.  Such was the option
in Aquarius Prolog, which defined the
BAM~\cite{DBLP:journals/computer/RoyD92}, significantly different from
the WAM, including several ``realistic'' low-level fine-granularity
instructions.  Likewise, native code generation within SICStus Prolog
led to the definition of an intermediate language, the SAM~(SICStus
Abstract Machine) which was translated into M68K assembler or,
alternatively, further compiled into yet another low level
representation (RISS) which was then mapped to a specific machine
language (Sparc or MIPS).
Neither survived: the Aquarius compiler remained unusably slow and
native code generation was dropped from SICStus because of its
maintenance requirements.

It can be argued that most efforts in designing and implementing
lower-level abstract machines for Prolog were targeting RISC
architectures.  For instance it used to be a challenge to effectively
use the fine-grained control of pipeline and instruction flow that was
typical of, say, MIPS or Sparc processors.  Nowadays, most available
microprocessors implement a common architecture (x86 or x86-64) but
specific hardware implementations have sufficiently differing pipeline
structures that it becomes very difficult to optimize for any one of
these.  Besides, dynamic instruction reordering also makes static
instruction scheduling a largely moot point.

Performance in modern architectures is heavily dependent on making
good use of the memory cache hierarchies; Prolog compiler writers
stand to gain a lot from making good use of cache organizations,
possibly more so than what can be bought by other optimization
techniques.  The problem is that there is a lot of variation across
systems that must be accounted for to extract optimal performance.

It turns out that the more sophisticated approaches to native code
generation for Prolog have somehow vanished in the long run, while
\gp{} remains up-to-date and has been ported to several low-level
architectures.
We feel we have achieved a good balance between simplicity,
maintainability and performance.
To pursue performance gains without sacrificing simplicity we are
investigating a replacement for the MA level in \gp{}: we are
presently evaluating tools such as LLVM~\cite{1281665} which can be
thought of as a typed, machine-independent assembly language.


\section{Finite-Domain Constraints}
\label{sec:fd-constraints}
The main extension built on top of {\wamcc} was arguably {\clpFD},
which added constraint solving over Finite Domains (FD).  {\gp}
compiles FD constraints in a way similar to its predecessor {\clpFD},
the approach being described in
\cite{DBLP:journals/jlp/CodognetD96,DBLP:conf/iclp/DiazC93}.  It is
based on a so-called ``RISC approach'' which consists of translating,
at compile-time, all complex user-constraints (e.g. disequations,
linear equations or inequations) into simple, primitive constraints
(the FD constraint system) which operate at a lower level and which
really embody the propagation mechanism for constraint solving.  We
shall first present the basic ideas of the FD constraint system and
then detail the extensions to this framework implemented in {\gp}.

The FD Constraint System was originally proposed by Pascal Van
Hentenryck in the concurrent constraint
setting~\cite{DBLP:journals/lncs/HentenryckSD94}, and an efficient
implementation in the {\clpFD} system is described in
\cite{DBLP:journals/jlp/CodognetD96,DBLP:conf/iclp/DiazC93}. FD is
based on a single {\em primitive constraint} with which complex
constraints are encoded: for example, constraints such as $X=Y$ or
$X\leq 2Y$ are {\em defined} by means of FD constraints, instead of
having to be explicitly built into the theory. Each constraint is made
of a set of propagation rules describing how the domain of each
variable is related to the domain of the other variables, i.e. rules
for describing node and arc consistency propagation (see for
instance~\cite{tsang} for more details on CSPs and consistency
algorithms.)

A {\em constraint} is a formula of the form {\Xinr} where $X$ is a
variable and $r$ is a range. A {\em range} in FD is a non empty finite
set of natural numbers. Intuitively, a constraint {\Xinr} enforces
that $X$ belongs to the range denoted by $r$. Such a range can be a
{\em constant range} (e.g. $1$..$10$) or an {\em indexical range},
when it contains one or more of the following:

\begin{itemize}

\item {\tt dom($Y$)} which represents the whole current domain of $Y$;

\item {\tt min($Y$)} which represents the minimal value of the current
domain of $Y$;

\item {\tt max($Y$)} which represents the maximal value of the current
domain of $Y$.

\item {\tt val($Y$)} which represents the final value $Y$ (i.e. the domain of
  $Y$ has been reduced to a singleton). A constraint involving such an
  indexical is delayed until $Y$ is determined.

\end{itemize}
\noindent
Obviously, when $Y$ is instantiated, all indexicals evaluate to its
value. When an {\Xinr} constraint uses an indexical term depending on
another variable $Y$ it becomes {\em store-sensitive} and must be
checked each time the domain of $Y$ is updated. This is how
consistency checking and domain reduction is achieved.

Complex constraints such as linear equations or inequations, as well
as symbolic constraints can be defined in terms of FD constraints,
see~\cite{DBLP:journals/jlp/CodognetD96} for more details.  For
instance, the constraint $X \leq Y$, is translated as
follows:\footnote{In this discussion, we're not using the \gp{}
  concrete syntax for constraint goals.}
\begin{center}
  X$ \leq $Y ~~ $\equiv$ ~~ X in 0..max(Y) ~~ $\wedge$~~
                 Y in min(X)..$\infty$          
\end{center}
\noindent
Notice that this translation also has an operational flavor and
specifies, for a given n-ary constraint, how the domain of a variable
may be updated in terms of the other variables. For example, consider
the FD constraint {\tt X in 0..max(Y)}: whenever the largest value of
the domain of \texttt{Y} changes (i.e. decreases), the domain of X
must be reduced. If, on the other hand, the domain of Y changes but
its largest value remains the same, then the domain of X does not
change. One can therefore consider those primitive {\tt X in r}
constraints as a low-level language in which to express the
propagation scheme. Indeed, it is possible to express in the
constraint definition (i.e.~the translation of a high-level user
constraint into a set of primitive constraints) the propagation scheme
chosen to solve the constraint: forward-checking, full or partial
look-ahead, according to the use of {\tt dom} or {\tt min}/{\tt max}
indexical terms.

\subsection{The Constraint Definition Language}

For {\gp}, we designed a specific language to define FD constraints
which is both flexible and powerful. The basic {\Xinr} is sufficient
to define simple arithmetic constraints but too restrictive to handle
constraints like $min(X,Y)=Z$ or reified constraints, both of which
need some form of delay mechanism. Another limitation is that it is
not possible to explicitly indicate the triggers for a particular
propagator: these are deduced from the indexical used in the {\Xinr}
primitives. The {\gp} constraint definition language, FD, has then
been designed to allow the user to define complex constraints and
proposes various constructs to overcome these limitations. FD programs
are compiled into C by the {\fd2c} translator. The resulting C program
is then compiled and the object fits into the compilation scheme shown
in Figure~\ref{compil-scheme}. We present the main features of the
constraint definition language by means of a few examples.

\subsubsection{Arithmetic Constraint Definition}

Consider a constraint $X+C=Y$ ($X$ and $Y$ are FD variables, $C$ is an
integer):

\begin{lstlisting}
x_plus_c_eq_y (fdv X, int C, fdv Y) {
  start X in min(Y) - C .. max(Y) - C        /* X = Y - C */
  start Y in min(X) + C .. max(X) + C        /* Y = X + C */
}
\end{lstlisting}
Constraints are defined in a C-like syntax. The head declares the name
of the constraint (\texttt{x\_plus\_c\_eq\_y}) and for each argument
its type (\texttt{fdv}, \texttt{int}) and its name. The keyword
\texttt{start} activates an {\Xinr} primitive. The first states that
the bounds of $X$ must be between $min(Y)-C$ and $max(Y)-C$.
Similarly, the second rule indicates how to update $Y$ from $X$.

Take a more complex example, which defines $min(X,A)=Z$ (where $X$ and
$Z$ are FD variables and $A$ an integer):

\begin{lstlisting}
min_x_a_eq_z (fdv X, int A, fdv Z) {
  start (c1) Z in Min(min(X),A)..max_integer /* Z >= min(X,A) */
  start (c2) Z in 0 .. max(X)                /* Z <= X */
  start (c3) X in min(Z) .. max_integer
  start      Z in 0 .. A                     /* Z <= A */

  wait_switch
     case A>max(Z)                           /* case A != Z */
        stop c1
        stop c2
        stop c3
        start Z in min(X) .. max(X)          /* Z = X */
        start X in min(Z) .. max(Z)
}
\end{lstlisting}
\noindent
The first {\Xinr} constraint uses a C macro \texttt{Min} to compute
the minimum of $min(X)$ and $A$. The keyword \texttt{max\_integer}
represents the greatest integer that an FD variable can take. Note the
use of the \texttt{wait\_switch} instruction to enforce $X=Z$ (and to
stop the constraints \texttt{c1}, \texttt{c2}, \texttt{c3}) as soon as
the case $A \neq Z$ is detected.

\subsubsection{Reified Constraint Definition}
\label{sec:clpfd-reified-bool}

The facility offered by the language to delay the activation of an
{\Xinr} constraint makes it possible to define reified constraints:
the basic idea of a reified constraint is to consider the truth value
of a constraint as a first-class object, which is given the form
(``reified'') of a boolean value. This allows the user to make
assumptions about the satisfiability of constraints in a given store
in order to conditionally require that other constraints be met. It is
feasible to use this mechanism, for instance, to define disjunctive
constraints, which can be very useful to model complex problems.

The following example illustrates how to define $X=C \Leftrightarrow
B$ where $X$ is an FD variable, $C$ an integer and $B$ a boolean
variable (i.e. an FD variable whose domain is $0$..$1$) which captures
the truth value of the constraint $X=C$.  The definition below waits
until either the truth of $X=C$ or the value of $B$ is known:

\begin{lstlisting}
truth_x_eq_c (fdv X,int C,fdv B) {
  wait_switch
     case max(B) == 0                  /* case B = 0  */
         start X in ~{ C }             /* X != C */
     case min(B) == 1                  /* case B = 1  */
          start X in { C }             /* X = C */
     case min(X) > C || max(X) < C     /* case X != C */
          start B in { 0 }             /* B = 0 */
     case min(X) == C && max(X) == C   /* case X = C  */
         start B in { 1 }              /* B = 1 */
}                                        
\end{lstlisting}
\noindent
Each constraint results in a C function returning a boolean
depending on the outcome of the addition of the constraint to the
store.  The link between Prolog and a constraint is done by the Prolog
built-in predicate \texttt{fd\_tell/1}.  For instance, to use the
previous constraint one could define the following predicate:
\begin{lstlisting}
'x=c <=> b'(X,C,B) :- 
    fd_tell(truth_x_eq_c(X,C,B)).
\end{lstlisting}

\subsubsection{Global Constraints}
\label{sec:clpfd-global-cstr}

Global constraints allow the user to specify patterns that are
frequently encountered in problems.  A global constraint can often be
decomposed into simple (local) constraints.  However the pruning
obtained with such a decomposition is less efficient than that
provided by specialized propagation algorithms.  The {\gp} constraint
language is not expressive enough to describe any filtering procedure
which has to be written in C. %
An API is provided to the C programmer for handling FD variables,
ranges, etc.  To simplify the interface between Prolog and C for
constraints, the FD language offers the following facilities:

\begin{itemize}

\item It handles lists of FD variables and/or integers (types \texttt{l\_fdv}
  and \texttt{l\_int}). At run-time, a corresponding Prolog list is
  expected and it is passed to the C code as a C array (of pointers to
  FD variables or of integers).

\item It implicitly wakes up the constraints suspended on indexicals
  occurring in the lists (but the user can define another triggering
  strategy).

\item It can invoke a user-defined C function to compute a range in a
  {\Xinr} primitive or outside any primitive.

\end{itemize}
\noindent
Consider the $element(I, L, V)$ constraint which says that the
$I$\textit{th} element of integer list $L$ is equal to $V$ ($I$ and
$V$ are FD variables).  It is defined as follows:

\begin{lstlisting}
pl_fd_element (fdv I, l_int L, fdv V) {
 start I in Pl_Fd_Element_I(L)
 start V in Pl_Fd_Element_I_To_V(dom(I), L)
 start I in Pl_Fd_Element_V_To_I(dom(V), L)
}
\end{lstlisting}
\noindent
The \emph{first constraint} is executed only once to set the initial
domain of $i$ to $1$..$\mathrm{length}(L)$. The \emph{second
  constraint} is woken up each time the domain of $I$ is modified, in
order to compute the new domain of $V$. To this effect, the C function
\texttt{Pl\_Fd\_Element\_I\_To\_V} is called. It mostly iterates over
each value $j$ from the domain of $I$, accumulating the
$j^\textit{th}$ element of the list $L$. The simplified C code of this
function looks like:

\begin{lstlisting}
void Pl_Fd_Element_I_To_V (Range *v, Range *i, WamWord *l) {
  int j;

  Vector_Allocate(v->vec);
  Pl_Vector_Empty(v->vec);

  VECTOR_BEGIN_ENUM(i->vec, j);
    Vector_Set_Value(v->vec, l[j]);
  VECTOR_END_ENUM;
}
\end{lstlisting}

\begin{description}
\small
\item[Line 1:] The function accepts \texttt{i} (the current domain of the
  variable $I$) and \texttt{l[]} (the array associated to the list of
  integers $L$) and computes \texttt{v}, the new domain of the variable $V$
  (this will be stored in the first argument of the function). Note:
  the \textit{tell} of $V in v$ is not done here but by the {\Xinr}
  primitive at the return of the function).

\item[Line 4-5:] A bit-vector is allocated and cleared (\texttt{v}).

\item[Line 7-9:] Using C macros, the values of the domain of $I$ are
  scanned. For each value \texttt{j}, the corresponding element in $L$
  (\texttt{l[j]}) is accumulated in \texttt{v}.

\end{description}

\noindent
Conversely, the \emph{third constraint} is triggered each time the
domain of $V$ is modified to compute the new domain of $I$. To this
end, the C function \texttt{Pl\_Fd\_Element\_V\_To\_I} iterates over
all values in $L$ which are also present in
$V$, accumulating their indexes.

{\gp} offers a variety of high-level constraints in the built-in
library. Low-level definitions of constraints as illustrated here are,
however, open to the expert programmer who needs to customize or
enrich the constraint solver for some practical application.  At the
moment the ultimate customization is achieved by writing C code. This
smoothly integrates into the native-compilation scheme adopted by
{\gp}.  We do plan, however, to extend the expressive power of the
language to be able to describe some global constraints without adding
any C code.

\subsection{Integrating Constraints into the WAM}
\label{sec:integr-constr-into}

We here recall the main points of the integration of FD constraints
into the WAM (see~\cite{DBLP:journals/jlp/CodognetD96} for more
detailed information).  To understand the necessary data structures
one needs to study the basic consistency procedure.  When a {\Xinr}
constraint is added, the range $r$ is evaluated and the domain of $X$
is updated accordingly (the new domain of $X$ being the intersection
between its current domain and $r$).  Once this is done, propagation
may occur: every constraint on $Y$ which depends on $X$
(e.g. \texttt{Y in min(X)+10..max(X)+10}) needs to be reevaluated.
Doing so will potentially modify $Y$ and all constraints depending on
$Y$ also need to be reconsidered.  The process finishes either when a
failure occurs (the new domain of a variable is empty) or when a
fix-point is reached (no more variables are modified).  In case of
failure, Prolog backtracking occurs.  It is then important to be able
to undo all modifications that have been done on the FD data
structures.

Adding constraints over finite domains to the \gp{} WAM required the
introduction of a new term type (FD Variable, with the \texttt{FDV}
tag) which, besides contributing to tag space depletion, needs to be
distinct from the regular \texttt{REF} term.  An \texttt{FDV} term has
2 distinct parts:

\begin{itemize}

\item Its domain: the set of allowable values, represented as the
  extrema of the containing interval or as discrete individual values,
  encoded as a bitmap, possibly multi-word. Using a bitmap greatly
  speeds up computation on sparse domains.

\item The dependencies: the set of constraints which depend on the
  variable, i.e.~those which need to be recomputed each time the
  variable is modified. In order to optimize the triggering of these
  constraints, several distinct chains are maintained (e.g. it is
  useless to re-execute a constraint depending on $min(X)$ when only
  $max(X)$ is changed).

\end{itemize}

Classically, a value-trail mechanism is used to save an FD variable
before its modification (domain and/or dependencies). On backtracking,
trailed values are used to restore the FD variable. In order to avoid
unnecessary trailings (for each FD variable, at most one trailing is
necessary per choice-point) a timestamp technique is used: a
sequential integer is used to number each choice-point and an FD
variable records the choice-point number associated to its last
trailing. This is important since FD variables are refined step by
step by the propagation algorithms which potentially compute several
intermediate domains before reaching the fix-point.

The two parts of an FD variable (domain and dependencies) are generally
not modified at the same time during the execution of a constraint
program. The dependency chains are created and updated when the
constraints are installed, typically at the start of program
execution, whereas the domains are more intensely modified later on,
for instance during the labeling phase which tries to find a solution
through backtracking. For this purpose, each part of an FD variable
(domain and dependencies) maintains its own independent timestamp. In
particular, when doing labeling, we only trail the domain of the FD
variables.

The other important data structure is the \emph{constraint frame},
which stores the information needed for constraint (re)evaluation.
For an {\Xinr} primitive we need:

\begin{itemize}

\item A pointer to the constrained variable $X$.

\item The address of the C function evaluating the range $r$ (this is
  produced by \texttt{fd2c} from the definition written in the
  constraint definition language).

\item A pointer to the environment in which the function evaluating
  $r$ executes: basically the function parameters, built by the
  constraint installation code.

\end{itemize}
\noindent
We chose %
a dedicated stack in which to store all these data
structures, called the \emph{constraint stack}.  As for other Prolog
data strucutures, the stack is used in backtracking: the top of the
constraint stack is saved in choice-points and restored when
backtracking occurs.  This was not the case in {\clpFD} where all FD
data structures were located in the heap.  In the tests we conducted,
the performance impact of having a constraint stack was negligible on
programs which did not use FD.

From the above propagation algorithm it appears that the evaluation of
a constraint leads to the reevaluation of other constraints.
Theoretically, the order in which constraints are woken up is not
relevant (since the process stops when the fix-point is reached). The
easiest way to implement this consists of a depth-first evaluation
(recursively calling each constraint depending on the variable which
has just been updated). However, this blind recursive descent is not
efficient in practice and misses some important optimizations. It is
thus better to explicitly handle a queue of constraints. A first
optimization consists of considering a queue of variables instead of a
queue of constraints. When a constraint needs to be reevaluated, it is
as consequence of the modification of some FD variable. It is easier
to record just this modified variable (a pointer) than to copy in the
queue all depending constraints. In {\gp} we go even further: the
queue is not separately represented: instead, all FD variables present
in the queue are linked together.  To this end, an FD variable (see
above) includes a third part which is devoted to the queue.  It
consists of:

\begin{itemize}
\item A link to the next enqueued variable (linked-list).

\item A mask describing which dependencies need to be reconsidered (to avoid
  useless reevaluations).

\item A time-stamp to know whether a variable is already present in
  the queue. There is a general counter which is incremented each time
  the (above) propagation procedure is run. When a variable is
  modified, if its time-stamp is different from the counter then the
  variable is not yet in the queue (it is then linked), otherwise only
  its mask of dependencies is updated.

\end{itemize}

Note that our choice for the representation of the queue associated
with the time-stamp technique described above results in an
optimization: the constraints depending on one variable are only
present once in the queue. On a set of benchmarks, this optimization
saves an average 17\% of the execution time (it is particularly
effective on arithmetic constraints) with no overhead. Detecting this
case with a separate queue would be much more time-consuming.

Another optimization which works well in practice for many constraints is
that an {\Xinr} primitive does not need to be evaluated if $X$ has been
instantiated before the start of the propagation procedure. This can be
detected reusing the same counter described above. On some examples this
optimizations saves up to 72\% of the execution time (in particular when many
disequalities are involved).

\BL

We have shown that {\gp} smoothly integrates an efficient FD solver,
proposing a simple yet powerful language in which to describe
high-level constraints and propagators.  Those constraints are compiled
down to C code, which in turn is integrated into the \gp{} executable
build flow of figure~\ref{compil-scheme}. More constraints can then
easily be added thanks to the description language and if needed with
the help of dedicated user-defined C functions. The compilation of
high-level constraints is based on a limited set of primitives which
are well optimized. These optimizations are ``general'' (vs. ``ad-hoc
optimizations'' of black-box solvers). So all high-level constraints
can benefit from them.



\section{\gp{} and the Prolog Standard}
\label{sec:gp-and-standards}

From the outset, \gp{} has aimed to comply with common practice in
Prolog implementations, while retaining its characteristic
architectural organization: to fit into a regular native code compiler
system, in which executables are produced by linking object modules.

\gp{} was developed at the same time as the ISO Core~1
standard~\cite{iso-part1}, which led us to take the standard proposal
into account from the outset.  \gp{} therefore became the first Prolog
system to closely comply with the ISO standard.  This meant supplying
not only the standard built-in predicates but also the related error
behaviours (e.g.~exceptions), the logical database update view for
dynamic predicates, meta-calls (the ISO standard requires a term to be
transformed into a goal before execution), directives, etc.
\footnote{It is worth
  noting that \gp{} has kept up with the proposed revisions to the
  standard, including features such as \texttt{call/N}, conditional
  compilation directives and evaluable functors, among others.}

We took compatibility one step further by providing a classical Prolog
top-level interpreter, with all the expected facilities operational,
including goal execution, source display (the \texttt{listing/0}
predicate), a trace-mode 4-port debugger, program consult and
reconsult, Prolog state manipulation operations (character
classification, operator definitions, etc.)  We do think that a
top-level interpreter is a primitive form of Integrated Development
Environment (IDE): it makes historic sense, but it would be better to
integrate stripped-down compiler-like tools into a graphical IDE such
as Eclipse, NetBeans or Xcode by means of a plug-in.

It can be argued that the DEC-10 Prolog compiler was influential in
many ways and some aspects of its design persist in today's Prolog
systems.  Its operating environment set a model which would be
emulated by most Prolog systems which came thereafter: the interactive
top-level with a ``workspace'' concept, which contains the whole
of the program, all seamlessly integrated, regardless of the
representation used for Prolog code: clausal form suitable for a
meta-interpreter, lower-level instructions adapted to a byte-code
interpreter or even executable native code.

This model holds, among others, for DEC-10 Prolog, C-Prolog, Quintus,
SICStus, MU-Prolog and Nu-Prolog, YAP, XSB, SWI, Ciao, BinProlog,
ECLiPSe and B-Prolog.  With such a heritage, \gp{} was almost
compelled to follow suit and establish itself around the concept of a
top-level interpreter managing goals executed in the context of a
dynamically-adjustable workspace, comprising all the Prolog modules
and equipped with a 4-port debugger, familiar to Prolog programmers,
although not strictly part of the language.

The functional enrichment of Prolog systems, and in particular those
features that stem from the language's meta-programming capabilities,
went forth basically unaware of the distinction between ``compile
time'' and ``execution time'' environments.  This mingling is such
that executing one query may influence a subsequent consult operation,
in ways that may completely alter its semantics (for instance,
operator definitions.)

For a system like \gp{}, which does independent static compilation to
native code and clearly separates the compilation from the execution
environments, providing a top-level interpreter similar to other
systems was a challenge which required a large development effort.
This requirement prompted the addition of a bytecode emulator to
\gp{}, similar to what is done in other Prolog implementations, to
provide a not-too-inefficient means of implementing the
culturally-accepted development cycle for Prolog programs:
edit/\-reconsult/\-run.  The grief over compile-time vs.~runtime
environments is not confined to \gp{} though: this is a prominent
issue in all systems that do static analysis or program
transformation, such as mode or type analyzers or even simple
pretty-printers.

We feel that the ISO standard missed a good opportunity to disentangle
this situation and separate compile from execution environments.  In
particular:

\begin{enumerate}

\item The \texttt{:- initialization} directive was meant for an
  interpreted environment, where one expects it to have an immediate
  effect on the rest of the program, whether it is simply being
  compiled or actually being loaded.  The semantics of this directive
  are unclear when the driving goal is, for instance, something like
  \texttt{consult([f1, f2])}, in which the initialization directives
  from \texttt{f1} may influence the loading of \texttt{f2}.  A
  possible way around this issue is to separate the execution of the
  initialization directives from the loading of the modules: \gp{}
  only executes the initializers once all modules have been loaded.
  The execution order is, in terms of the ISO standard,
  ``implementation dependent.''

\item Another ISO directive which causes grief is
  \texttt{multifile/1}: one problem is the order in which the multiple
  batches of clauses get collected.  This is not an issue in an
  interpreted environment, in which the loading is explicitly
  controlled by the programmer whereas in a statically compiled set of
  Prolog files the order is largely unpredictable, because it is left
  to the linker's criteria.

\item One unfortunate feature of the Prolog language, legacy of the
  interpreter tradition, is the lack of distinction between code and
  data-only (database) dynamic predicates.
  The ISO standard missed the opportunity to clearly distinguish
  between these two traditional uses for dynamic predicates:
  persistent data and dynamic code manipulation.

\end{enumerate}

\noindent
These were but a few of the difficulties which hit us when developing
\gp{}; nevertheless, we strived to
  provide a fair rendering of an expected set of built-in predicates.
  The ``standard'' Prolog library is nowhere near as complex as that
  of other languages so the extent of this requirement is limited.
We also
behave conservatively w.r.t.~extra-logical aspects of Prolog, such as
the handling of directives.


\section{Conclusions and Directions for Future Work}
\label{sec:concl}
We have presented the most significant aspects of the implementation
of \gp{} for which the key issues were simplicity, extensibility and
maintainability without sacrificing performance.  This led us to the
native-code generation approach which has been described in this
article. We applied the same requirements for the design and the
development of the finite domain constraint solver. One might say
that, overall, the \gp{} experience has been successful. The \gp{}
``family'', which includes \wamcc{} and \clpFD{} has been used in
teaching and as the basis of several extensions, most notably by other
research teams: this fulfills one of our design goals which was to
establish a system sufficiently simple for it to be easily extended by
other people.

Several architectural ports and some extensions have been provided by
the user community, as acknowledged in the \gp{} distributed
documentation.  Modified versions of \gp{} have been used for
prototyping systems, featuring module systems, threads, attributed
variables, CLP$(\cal R)$,
RDBMS integration, Java interfaces, a MacOSX IDE, to name but a
few.  The community-supplied extensions are referenced on the main
site at \url{http://www.gprolog.org/#contribs}.

In what concerns dissemination, the \gp{} distribution had been
downloaded well over 100000 times from the development FTP site, over
a period of four years.
We no longer keep statistics, as \gp{} is part of several Linux
distributions and there is no way to account for downloads from the
main GNU FTP site nor from other mirrors.

Performance-wise, \gp{} scores honorably, barely below
YAP~\cite{DBLP:journals/jucs/SilvaC06} which is continually being
tuned for performance.  We compared GNU Prolog 1.4.0 and YAP 5.1.3 on
64-bit Linux.  On the average, YAP is faster by factor of 1.3 with
peaks up to 2. However, on some benchmarks GNU Prolog can outperform
YAP by a factor of up to 1.4.  With respect to \wamcc{} we clearly
gained in usability, as a consequence of the more realistic compile
times.
\gp{} is currently being worked on in various directions, including:
\begin{itemize}
\item \textbf{Modules}: \gp{} initially did not implement any module
  system, staying within the bounds of ISO Prolog Core 1, awaiting the
  ISO Modules specification.  Reaching a consensus on modules took a
  long time and the resulting specification is still not very
  satisfactory.  We initially opted for the implementation of a
  cleaner alternative mechanism, Contextual Logic
  Programming~\cite{DBLP:conf/iclp/AbreuD03}.
  Nevertheless, as there is a clear need for an interoperable module
  system, we are finishing a minimal-functionality module system as
  part of the Prolog Commons initiative, which brings \gp{} at par
  with the other implementations.

\item \textbf{Other ISO Prolog Features}: ISO compliance has been
  foremost in the design and implementation of \gp{}; the work being
  carried out by the ISO standization committee is being actively
  followed.  For instance, one aspect that needs to be accounted for
  is the handling of Unicode characters.

\item \textbf{Attributed Variables}: even though \gp{} has a very
  efficient, convenient and easily extensible FD constraint solver, it
  makes sense to include other constraint domains.  Attributed
  variables are a mechanism which can be used to effectively implement
  constraints and propagation over other domains.

\item \textbf{Tabling}: using tabling allows one to write programs
  which are more expressive because the system takes care of memoizing
  for us.  More programs terminate which would otherwise loop and this
  can be a very effective programming device.  This extension to
  Prolog was introduced in XSB~\cite{rao97xsb} and has since been
  included in other systems, namely
  YAP~\cite{rocha2000yaptab,DBLP:journals/tplp/RochaSC05} and
  B-Prolog~\cite{DBLP:journals/tplp/ZhouSS08}.

\item \textbf{A Garbage Collector}: \gp{} has gotten by without GC.
  While reasonable for short-lived
  processes\footnote{
    \gp{} is very suitable for building small executables, with short
    load times: Unix systems share code sections and this means code
    lingers even when no processes are using it.  This is normally not
    the case with bytecodes kept in writeable, non-shared memory.} it
  is a limiting factor for larger executions.

\item \textbf{Improved Compiler}: the \gp{} Prolog-to-WAM compiler is
  rather simple.
  This is an obvious area for improvements.

\item \textbf{Compilation Pipeline}: because it is made up of a
  succession of filters, \gp{} is amenable to the substitution of some
  of these: we are presently working on a few, for instance one which
  manages EAM-style executions from the WAM code.  A longer-term goal
  is to rework the \gp{} back-end and improve its integration into an
  existing compiler scheme: LLVM~\cite{1281665} is an interesting
  target, as it is essentially a machine-independent, typed assembly
  language which could take over the MA and Assembly language steps.

\end{itemize}

\paragraph{Acknowledgements:}
The authors wish to acknowledge the anonymous reviewers who read early
versions of this text and whose criticisms contributed to
significantly improve its form and substance.


\bibliography{gprolog-tplp}

\end{document}